# Valley vortex states and degeneracy lifting via photonic higher-band excitation

**Daohong Song[1,2], Daniel Leykam [3], Jing Su[1], Xiuying Liu[1], Liqin Tang[1], Sheng Liu[4], Jianlin Zhao[4], Nikolaos K. Efremidis[5], Jingjun Xu[1,2], and Zhigang Chen[1,2,6]**

[1] *The MOE Key Laboratory of Weak-Light Nonlinear Photonics, and TEDA Applied Physics Institute and School of Physics, Nankai University, Tianjin 300457, China*

*2 Collaborative Innovation Center of Extreme Optics, Shanxi University, Taiyuan, Shanxi 030006, China*

[3]*Center for Theoretical Physics of Complex Systems, Institute for Basic Science, Daejeon 34126, Republic of Korea*

[4]*The Key Laboratory of Space Applied Physics and Chemistry, Ministry of Education and School of Science, Northwestern Polytechnical University, Xi'an 710072, China*

[5]*Department of Mathematics and Applied Mathematics, University of Crete, 70013 Heraklion, Crete, Greece*

[6]*Department of Physics and Astronomy, San Francisco State University, San Francisco, CA 94132*

*Corresponding author: dleykam@ibs.re.kr, jjxu@nankai.edu.cn, zgchen@nankai.edu.cn*

**Abstract:** We demonstrate valley-dependent vortex generation in a photonic graphene. Without breaking the inversion symmetry, excitation of two equivalent valleys leads to formation of an optical vortex upon Bragg-reflection to the third valley, with its chirality determined by the valley degree of freedom. Vortex-antivortex pairs with valley-dependent topological charge flipping are also observed and corroborated by numerical simulations. Furthermore, we develop a three-band effective Hamiltonian model to describe the dynamics of the coupled valleys, and find that the commonly used two-band model is not sufficient to explain the observed vortex degeneracy lifting. Such valley-polarized vortex states arise from high-band excitation without inversion symmetry breaking or synthetic-field-induced gap opening. Our results from a photonic setting may provide insight for the study of valley contrasting and Berry-phase mediated topological phenomena in other systems.

Valley pseudospin (VSP), manifesting the degenerate energy extrema of the bands in momentum space, is an intriguing fundamental concept that has stimulated tremendous interest. In recent years, light–valley interaction in two-dimensional (2D) materials has attracted a great deal of multidisciplinary interest in condensed matter physics and optoelectronics, largely due to its importance in fundamental physics and potential applications [1-3]. In particular, the VSP (or *valley degree of freedom* (VDF)) can be exploited to encode information for electrons just as the spin in spintronics, leading to a revitalizing field of "valleytronics" where a crucial step is to obtain pure valley-polarized states [4,5]. In 2D materials such as $MoS_2$, such valley polarization electron states can be obtained by applying various external fields, which leads to different behavior due to the opposite Berry curvature of the two inequivalent valleys [2]. For example, when applying an electric field, the valley states of the electron can be spatially separated due to the valley Hall effect. Interestingly, the concept of the VDF and associated valley vortex states has been extended to classical wave systems involving artificial honeycomb lattices, ranging from optics to acoustics [6-13]. Such a utilization of the unique topological features of the valley states provides a new way to manipulate waves with robust transport properties. In these endeavors, various methods to achieve non-topological bulk valley transport as well as topological valley transport in domain walls have been demonstrated in various systems [10-17]. Thus far, most of studies on valley-polarized states have been realized in honeycomb lattices (HCLs) with inversion symmetry breaking or with other synthetic gauge fields that open the gap at the Dirac point.

Photonic graphene, an HCL composed of evanescently coupled waveguides arranged in a honeycomb geometry [18-22], serves as a compelling platform to emulate graphene and topological physics. The advantage of using a photonic HCL system not only lies in its controllable structure parameters, but also in its Bloch modes with desired momentum that can be selectively excited and directly measured in both intensity and phase, enabling the exploration of fundamental phenomena which are otherwise inaccessible in real graphene materials. Indeed, a variety of intriguing phenomena has been observed in photonic graphene in recent years, including for example unconventional edge states, pseudospin-mediated vortex states, Aharonov-Bohm-like interference, valley Landau-Zener-Bloch oscillations, photonic topological insulator and topological valley Hall states [12,23-28].

In this work, we demonstrate the generation of valley contrasting vortices and vortex-pairs in optically induced HCL with preserving lattice inversion symmetry [29]. Specifically, when two interfering Gaussian-like beams are mapped onto the same sublattice (in real space) but selectively excite two equivalent valleys (either *K* or *K'* in momentum space), a singly-charged vortex emerges at the HCL output with its vortex chirality determined by the valley selection due to the time reversal symmetry. Furthermore, valley-dependent vortex-antivortex pairs are also observed when the two sublattices are equally excited. Numerical results obtained from the paraxial Schrödinger-type equation are in excellent agreement with experimental observations, uncovering topological charge flipping from decomposed two spinor components. In contradistinction to previously observed pseudospin states based on the *sublattice degree of freedom* (SDF) [26,30], charge flipping based on the VDF occurs in the same spinor component. Importantly, we develop an effective Hamiltonian three-band model to describe the coupled valley dynamics, and show that the presence of a third gapped band is essential to explain the observed vortex-pair states, although the gap itself remains closed at the Dirac points where a singular Berry curvature and pseudospin winding number is expected [31,32]. Such vortex degeneracy lifting without inversion symmetry breaking or synthetic-field-induced gap opening may lead to new understanding of VSP-mediated topological phenomena in other systems.

The dynamics of a probe beam propagating through the photonic HCL can be simulated by the following paraxial Schrödinger-type equation [33]:

$$i\frac{\partial\psi(x,y,z)}{\partial z}=-\frac{1}{2k_0}\nabla^2\psi(x,y,z)-\frac{k_0\Delta n(x,y)}{n_0}\psi(x,y,z)\equiv H_0\psi, \tag{1}$$

where $\psi$ is the field envelope of the probe beam, ($x$, $y$) are the transverse coordinates, $z$ is the longitudinal propagation distance, $k_0$ is the wavenumber, $n_0$ is the refractive index of the nonlinear medium, and $\Delta n(\mathrm{x,y})$ is the induced index change forming the HCL. In Eq. (1), $H_0$ is the continuous Hamiltonian of the system, whose eigenvalues are the wavevectors along the *z*-direction (i.e., the propagation constant). The HCL is composed of two triangular sublattices *A* and *B* in real space (Fig.1(a)), while in momentum space the first two Floquet-Bloch bands intersect at the Dirac points located at the corners of the first Brillouin zone (BZ), namely the *K* and *K'* valleys (see Fig.1(b)), where the dispersion is linear [34]. Around the Dirac points, the Berry flux is $\pi$ and $-\pi$ for the *K* and *K'* valleys, respectively, as measured in real graphene [35] or an artificial graphene system [36]. In order to detect the valley-contrasting phenomena, it is often necessary to break the lattice inversion symmetry, opening a gap at the Dirac points to examine the valley-dependent transport based on the Berry curvature [37]. In contrast, in what follows, we study the valley vortex states in a photonic HCL without breaking the inversion symmetry by directly measuring the phase structure of the valley states.

First, we numerically examine the valley-dependent vortex generation and topological charge flipping when inequivalent valleys of the HCL are selectively excited. Typical simulation results are shown in Fig. 1. The HCL, shown in Fig. 1(a), is uniform and thus has the inversion symmetry, hosting gapless touching of the two bands at the Dirac points [18,19]. The probe beam, depicted in Fig. 1(b), is constructed by interfering two broad Gaussian beams, momentum-matched to two Dirac points in the same (either *K* or *K'*) valleys [Fig.1 (c)]. Here, we focus on the VDF in momentum space, while keeping the same sublattice excitation in real space (e.g., sublattice *A*). After propagating through the HCL, the output probe beam exhibits asymmetrical conical diffraction. In particular, by comparing the output intensity patterns when exciting the K or the K' valleys we see that they are related by reflection symmetry. Although the probe beam excites only two Dirac points initially, the spectrum at the third equivalent Dirac point emerges at output due to Bragg reflection (see the blue dashed circles in Fig.1(e)). Interestingly such a Bragg-reflected component (extracted in Fourier space) contains a singly-charged vortex with opposite chirality at $K_3$ and $K_3$', as can be seen clearly from the phase structure of the valley states (Fig.1(f)).

Next, we present experimental results of such valley-dependent vortex states. The setup is similar to that used in our previous work on sublattice-mediated vortex generation [26,38] except that now the excitation conditions are different. The HCL is optically induced in a 10mm-long biased photorefractive SBN crystal illuminated with a triangular lattice beam. When applying a negative voltage (about 1.4 kV/cm) against the crystalline *c*-axis, the lattice beam experiences a self-defocusing nonlinearity which transforms the triangular intensity pattern into the HCL index potential [18,23,39]. For this study, the lattice spacing is about 16μm as shown in Fig. 2(a). The *k*-space spectrum of the lattice beam is shown in Fig. 2(b), where *K* and *K*' valleys are located at the corners of the white dashed lines. The probe beam is constructed by interfering two broad Gaussian beams with their launching angles aligned to match two *K* (or two *K'*) valleys, as illustrated in Fig. 1(c).

Typical experimental results from the valley-dependent excitations are shown in Figs. 2(c)-2(d). As in the simulation results of Fig. 1(d), the output patterns are asymmetric with a mirror reflection symmetry for different valley excitations. The Bragg-reflected components at the $K_3$ or $K_3'$ valley are marked with blue dashed circles in the bottom subpanels of Fig. 2(d). In order to measure the phase of this newly generated component, an inclined reference beam is introduced to interfere with the far-field intensity pattern extracted from the $K_3$ or $K_3'$ valley. The resulting interferograms clearly display a single fork bifurcation towards opposite directions as shown in Fig.2(e), indicating opposite topological charges. These results agree well with the simulation results shown in Fig.1(f). Note that the probe beam excites the same sublattice *A* for both cases, thus the vortex charge flipping observed here is not due to the SDF [26] but rather the VDF. Interestingly, momentum-space vortices have also been observed in periodic plasmonic structures arising from winding of polarization vectors [40].

In addition to above singly-charge vortex, a valley-dependent vortex-antivortex pair and associated topological charge flipping is also observed. Experimentally this is realized when both sublattices are equally excited, by launching the probe beam half-way between the two sublattices (see the insert in Fig.3(a)). The results for such a vortex pair generation are summarized in Fig.3, where Fig.3(a) shows the overall output intensity patterns, whereas Fig.3(b) shows the far-field intensity patterns from the newly generated $K_3$ (top panel) or $K_3$' (bottom panel) component. Although these intensity patterns are not dramatically different, their phase structure unveils new information. The interferograms from the Bragg-reflected components exhibit clearly two (instead of one) separate fringe bifurcations towards opposite directions as shown in Fig.3(c). Moreover, the chirality of the pairing vortices is reversed when the excitation switches from the *K* valleys to the *K'* valleys. These results demonstrate clearly valley-dependent charge flipping of the vortex pair when both sublattices are equally excited, as observed also in numerical simulations (Fig.3(d)). The difference in vortex position between experiments and simulations is attributed to the lattice inhomogeneity and/or imperfect excitation conditions in experiment.

When modeling the SDF, it is often helpful to decompose the optical field into the two sublattices or spinor components $\psi = (\psi_A, \psi_B)$ [26]. In a similar fashion, the valley vortex states are usually written in the momentum space spinor form $\phi = (\phi_A, \phi_B)$, where $\phi_A$ and $\phi_B$ are the Fourier transformation of $\psi_A$ and $\psi_B$, respectively. By solving Eq. (1), we obtain the intensity and phase of the two spinor components numerically for the *K* valley vortex states, as shown in Fig.4, where the two components have similar asymmetric intensity patterns but different phase structures. The top panels are derived from on-site excitation, where $\phi_A$ component has a flat phase structure [Fig.4 (b)] but $\phi_B$ component manifests a singly-charged vortex [Fig.4 (d)]. Thus, the superposition of two spinor components leads to a singly-charged vortex at the $K_3$ valley, in agreement with Fig. 2(e). On the other hand, the bottom panels are from inter-site excitation for vortex-antivortex pair generation. In this case, each spinor component has a singly-charged vortex but with opposite chirality [Figs. 4(b) and 4(d)], and thus the superposition leads to a vortex pair, as observed in Fig.3(c). For the *K'* valley vortex states, the chirality of all spinor vortices are reversed due to the time reversal symmetry. Note that, for sublattice-dependent vortex states, the topological charge flipping occurs in two different spinor components while only one sublattice is excited [26], but for valley-dependent vortex states, it occurs in the same sublattice spinor component. Interestingly, from the results shown in Fig. 3 the valley vortex pair is generated with the two vortices well separated at the output, whereas from the sublattice decomposition analysis the vortices in the two components seem to overlap, suggesting the vortices should be degenerate. The vortex degeneracy can be lifted if the HCL is deformed [27,28], but in our case it is not. The underlying physics of the vortex pair generation certainly merits further investigation.

Thus, we develop a theoretical model to directly analyze the wave dynamics in momentum space. As the HCL is established by self-defocusing nonlinearity in experiment, the index lattice potential can be written as [18,21] $2k_0^2\Delta n/n_0 = -V(e^{i\boldsymbol{G}_1\cdot\boldsymbol{r}} + e^{i\boldsymbol{G}_2\cdot\boldsymbol{r}} + e^{i(\boldsymbol{G}_1+\boldsymbol{G}_2)\cdot\boldsymbol{r}} + \text{c.c.})$, where $\boldsymbol{G}_{1,2} = \frac{2\pi}{a}(\frac{1}{\sqrt{3}}, \pm 1)$ and $\boldsymbol{G}_3 \equiv \boldsymbol{G}_1 + \boldsymbol{G}_2$ are the reciprocal lattice vectors. Introducing the normalized propagation distance $\tilde{z} = 2k_0 z$ and expanding $\psi(\boldsymbol{r}) = \int d\boldsymbol{p}\psi(\boldsymbol{p})\,\mathrm{e}^{i\boldsymbol{p}\cdot\boldsymbol{r}}$, the paraxial equation (Eq.(1)) can be Fourier transformed to obtain

$$i\partial_{\tilde{z}}\psi(\boldsymbol{p}) = |\boldsymbol{p}|^2\,\psi(\boldsymbol{p}) + V\sum_{j=1}^{3}\left[\psi(\boldsymbol{p}-\boldsymbol{G}_j) + \psi(\boldsymbol{p}+\boldsymbol{G}_j)\right] \quad , \tag{2}$$

where $\boldsymbol{p}$ is the wavepacket momentum, and $V$ is a constant. The first term on the right hand side forms a parabolic confining potential and accounts for the kinetic energy term, while the second term describes coupling to the six neighboring reciprocal lattice points. Thus the Fourier space propagation resembles the tight-binding dynamics of a hexagonal lattice with a super-imposed parabolic potential.

If the lattice potential is weak, the parabolic confining potential $|\boldsymbol{p}|^2$ dominates and thus suppresses the scattering to higher BZs. Therefore, we can restrict our attention to the beam dynamics within the 1st BZ, in particular the coupling between the three equivalent valleys $\boldsymbol{K}_{1,2,3}$, where $\boldsymbol{K}_1 = \frac{2\pi}{a}(-\frac{1}{\sqrt{3}}, -\frac{1}{3})$, $\boldsymbol{K}_2 = \boldsymbol{K}_1 + \boldsymbol{G}_1 + \boldsymbol{G}_2$, and $K_3 = K + G_1$ are the positions of the equivalent Dirac points [41]. Assuming small displacements from the Dirac points, $\boldsymbol{P} = \boldsymbol{K}_{1,2,3} + \boldsymbol{\delta p} = \boldsymbol{K}_{1,2,3} + \delta p(\cos\theta, \sin\theta)$, the dynamics of the three coupled valleys can be described by an effective Hamiltonian: $i\partial_z \Psi(\delta \boldsymbol{p}) = \hat{H}_{eff}\, \Psi(\delta \boldsymbol{p})$. After expending the kinetic energy term, the effective Hamiltonian can be written as:

$$\hat{H}_{eff}(\delta \boldsymbol{p}) = p_0^2 + \delta p^2 + \begin{pmatrix} -\frac{4\pi}{3}\delta p(\sqrt{3}\cos\theta + \sin\theta) & V & V \\ V & \frac{4\pi}{3}\delta p(\sqrt{3}\cos\theta - \sin\theta) & V \\ V & V & \frac{8\pi}{3}\delta p \sin\theta \end{pmatrix}, \quad (3)$$

which describes the dynamics in the coupled valleys $\boldsymbol{K}_{1,2,3}$. For *K’* valleys, the only difference is the sign in the $\sin\theta$ term, which corresponds to the *y*-component of the displacement from the valleys. At large momenta, this model exhibits trigonal warping of the Dirac cones, which can lift the valley degeneracy [42].

For this effective model, the Dirac point lies at $\delta p = 0$, where the eigenvalues of the Hamiltonian matrix $\hat{H}_{eff}$ are –*V*, -*V*, 2*V* with corresponding eigenvectors $|u_1\rangle = (-1,1,0)$, $|u_2\rangle = (-1,0,1)$, and $|u_3\rangle = (1,1,1)$, where $|u_{1,2}\rangle$ are the degenerate Dirac point eigenstates, while $|u_3\rangle = (1,1,1)$ is the eigenstate of a third (gapped) band. Roughly speaking, this third band describes inter-site excitations between waveguides/potential minima (i.e., the anti-guiding modes), which is the case for the excitation of valley vortex pairs as shown in Fig. 3.

We can construct pseudospin eigenstates as rotationally-symmetric superposition of $|u_{1,2}\rangle$, i.e. $|u_\pm\rangle = (1, e^{\pm i2\pi/3}, e^{\mp i2\pi/3})$. To verify that this is the correct form of the pseudospin eigenstates, we rewrite $\hat{H}_{eff}$ in the basis formed by $(|u_+\rangle, |u_-\rangle, |u_3\rangle)$:

$$\hat{H}_{spinbasis} = p_0^2 + \delta p^2 - \begin{pmatrix} V & \frac{2\pi}{3}(\sqrt{3}+i)\,\delta p e^{-i\theta} & \frac{2\pi}{3}(\sqrt{3}-i)\,\delta p e^{i\theta} \\ \frac{2\pi}{3}(\sqrt{3}-i)\,\delta p e^{i\theta} & V & \frac{2\pi}{3}(\sqrt{3}+i)\,\delta p e^{-i\theta} \\ \frac{2\pi}{3}(\sqrt{3}+i)\,\delta p e^{-i\theta} & \frac{2\pi}{3}(\sqrt{3}-i)\,\delta p e^{i\theta} & -2V \end{pmatrix}. \quad (4)$$

If we assume the third band is not excited and can be neglected (*i.e.* in the tight-binding limit of a deep lattice potential - not really the case for our experiment), then the above equation turns into the familiar massless two-band Dirac Hamiltonian.

To compare with the valley vortex experiment, we excite two equivalent valleys with a tunable relative phase between them, *i.e.* the initial state is $\Psi = e^{-\delta p^2 \omega^2}(1, e^{i\varphi}, 0)$, where $\omega$ is the beam width and $\varphi$ determines the position of the peak amplitude associated with the beam in real space. The beam has a peak at the space between waveguides when $\varphi = 0$, π (as in the case of inter-site excitation in experiment), but on one of the sublattices when $\varphi = \pm 2\pi/3$ (as in the case of on-site excitation in experiment). Note that the two beam excitation always excites the third band $|u_3\rangle$, except for the special case $\varphi = \pi$ .

We excite only one sublattice by setting $\varphi = 2\pi/3$ and compute the field scattered into the third (initially unexcited) valley by solving Eq. (4). The calculated phase and intensity is shown in Fig. 5, in good agreement with the experimental results. Swapping the valley index (while keeping the sublattice fixed) flips the topological charge, leading to the generation of a vortex-antivortex pair. The vortex degeneracy is lifted by the emerging of the third band mode due to the off-site excitation, in addition to the modes at the vicinity of Dirac points. So our effective model can explain the experimental observations and capture the essential physics behind the vortex generation. To apply a two band approximation, one must project the initial state onto the Dirac cone modes of $\hat{H}_{eff}$ . However, doing so results in a nonzero input field amplitude at the 3$^{rd}$ Dirac point. Basically, the input wavepacket momentum $\boldsymbol{p}$ and the propagation constant/energy eigenvalue $\boldsymbol{k}_z$ do not commute, i.e. plane wave/Gaussian inputs at one or two $K$ points are not Bloch wave eigenstates, which excite all three *K* points.

In conclusion, we have experimentally demonstrated valley-dependent vortex generation and topological charge flipping in optically induced HCL without breaking the lattice inversion symmetry. In contrast to the SDF, the VDF flips the topological charge in the same spinor component. Moreover, the valley vortex pair generation is achieved due to the lifting of the spinor vortex degeneracy arising from excitation of the higher band modes, as explained by the momentum space three-band model. Our results show clearly that valley-polarized states can be realized in HCL without inversion symmetry breaking or other synthetic gauge fields to open the gap at the Dirac point, and that the commonly used two-band model is not always sufficient in the study of graphene-related physical phenomena. Thus, we believe our work should provide insight in the areas of valley physics in both condensed matter and artificial graphene systems. In addition to the sublattice pseudospin, the VSP adds a new degree of freedom and provides unique features for manipulation of light propagation in photonic structures.

**Acknowledgement**

This research is supported by The National key R&D Program of China under Grant (No. 2017YFA0303800, No. 2017YFA0305100), the National Natural Science Foundation (11674180，11304165), PCSIRT（IRT0149, the 111 Project (No. B07013) in China, and the Institute for Basic Science in Korea (IBS-R024-Y1).

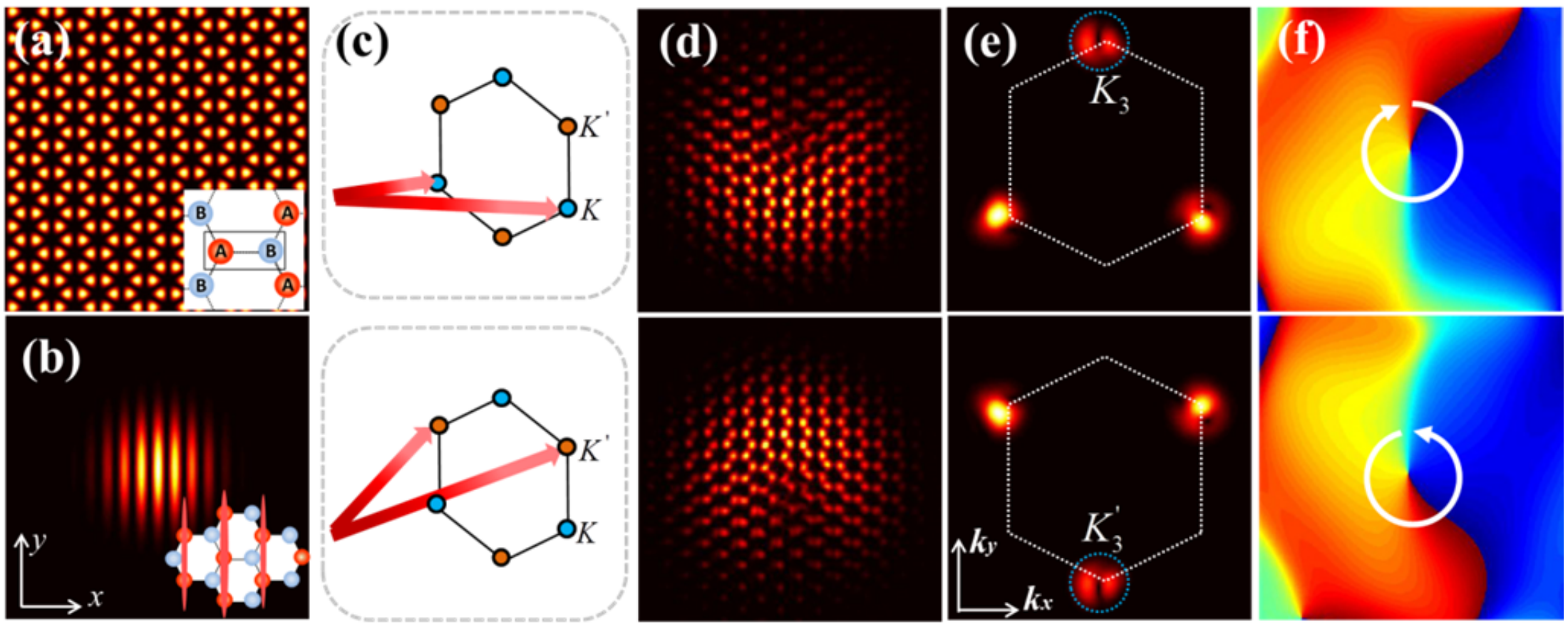

**Fig.1:** Numerical simulation of pseudospin vortex states from valley-selective excitation. (a) The HCL composed of induced waveguides, where the inset illustrates sublattices *A* and *B* marked with red and blue dots, respectively. (b) The interfering probe beams used for lattice excitation, where the inset illustrates the excitation condition in real space. (c) Illustration of the excitation condition when the two *K* (top row) or *K'* (bottom row) valleys are selectively excited in momentum space. (d) Output intensity pattern of the probe beams. (e) Corresponding spectrum, where blue dashed circle marks the Bragg-reflected component, and white dashed hexagon depicts the first BZ. (f) Phase structure of the Bragg-reflected component.

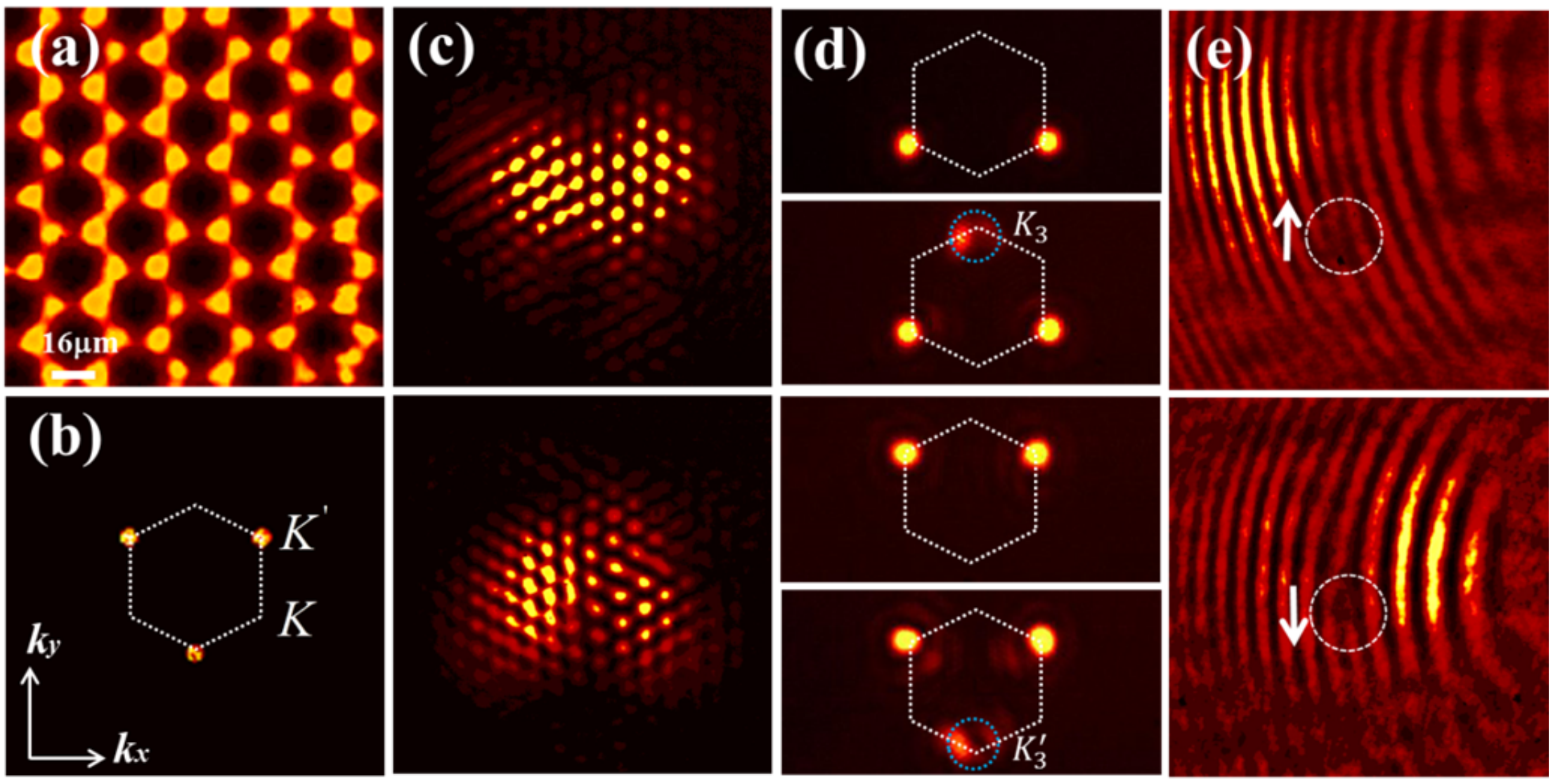

**Fig.2** Experimental demonstration corresponding to Fig. 1. (a) The HCL established by optical induction. (b) The *k*-space spectrum of the lattice beam, where *K* and *K*' valleys are marked. (c-e) Vortex generation when two *K* (top row) or *K*' (bottom row) valleys are selectively excited. (c) Output intensity pattern. (d) Input (top subpanel) and output (bottom subpanel) spectra, where blue dashed circles mark the newly generated spectral component at the third valley. (e) Interferograms from the Bragg-reflected components, where the vortex position/chirality is marked.

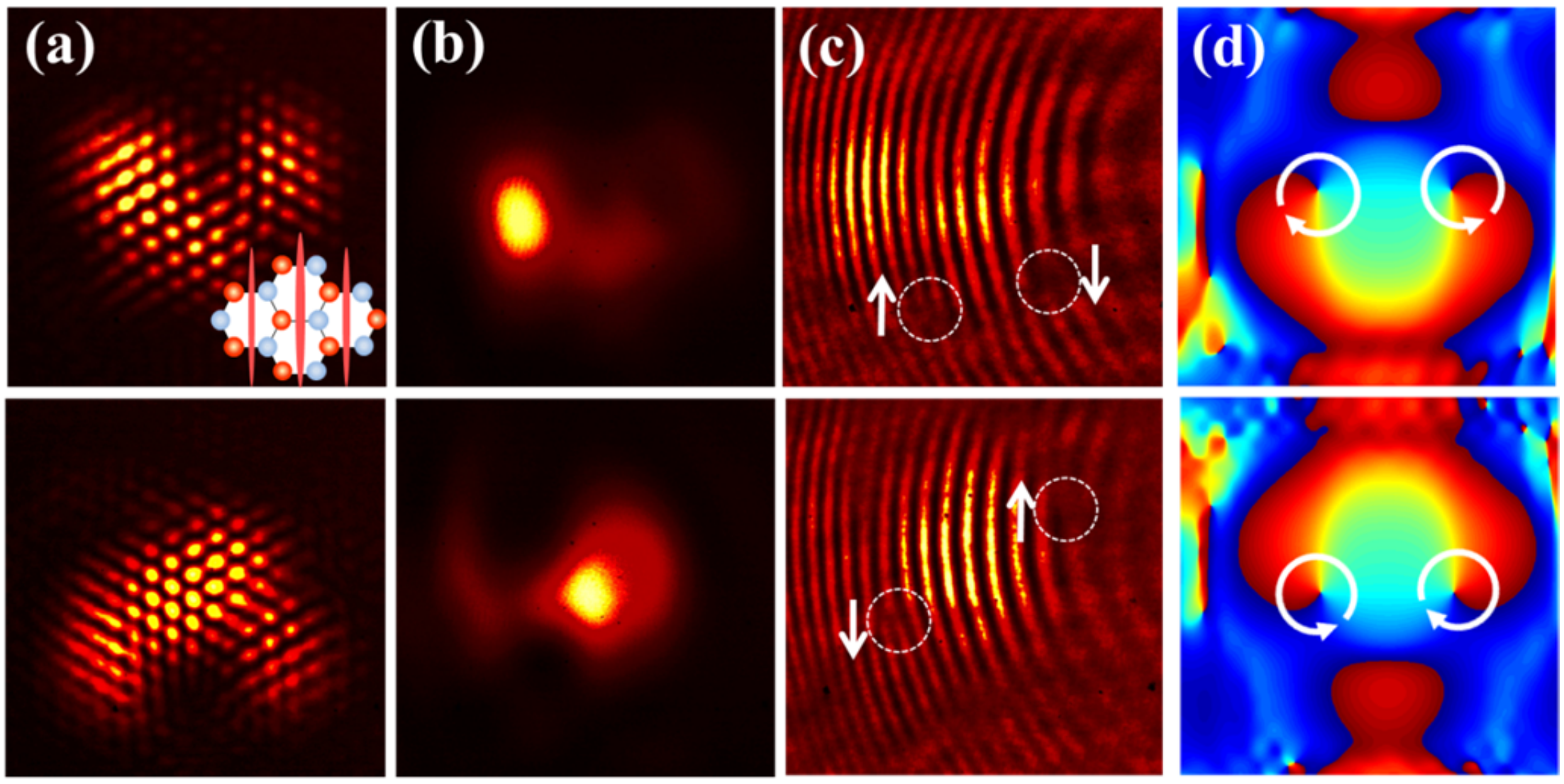

**Fig.3:** Demonstration of valley-dependent vortex pair generation. The excitation scheme is the same as in Fig. 1 for *K* (top row) or *K*' (bottom row) valleys, except that the probe beam is shifted to inter-site position (see inset in (a)). (a) Output intensity patterns. (b) Far-field patterns of Bragg-reflected component from the third valley. (c) Interferograms of (b) with an inclined plane wave, where the vortex pair is marked. (d) Numerically calculated phase of the Bragg-reflected component showing the vortex pair of opposite chirality.

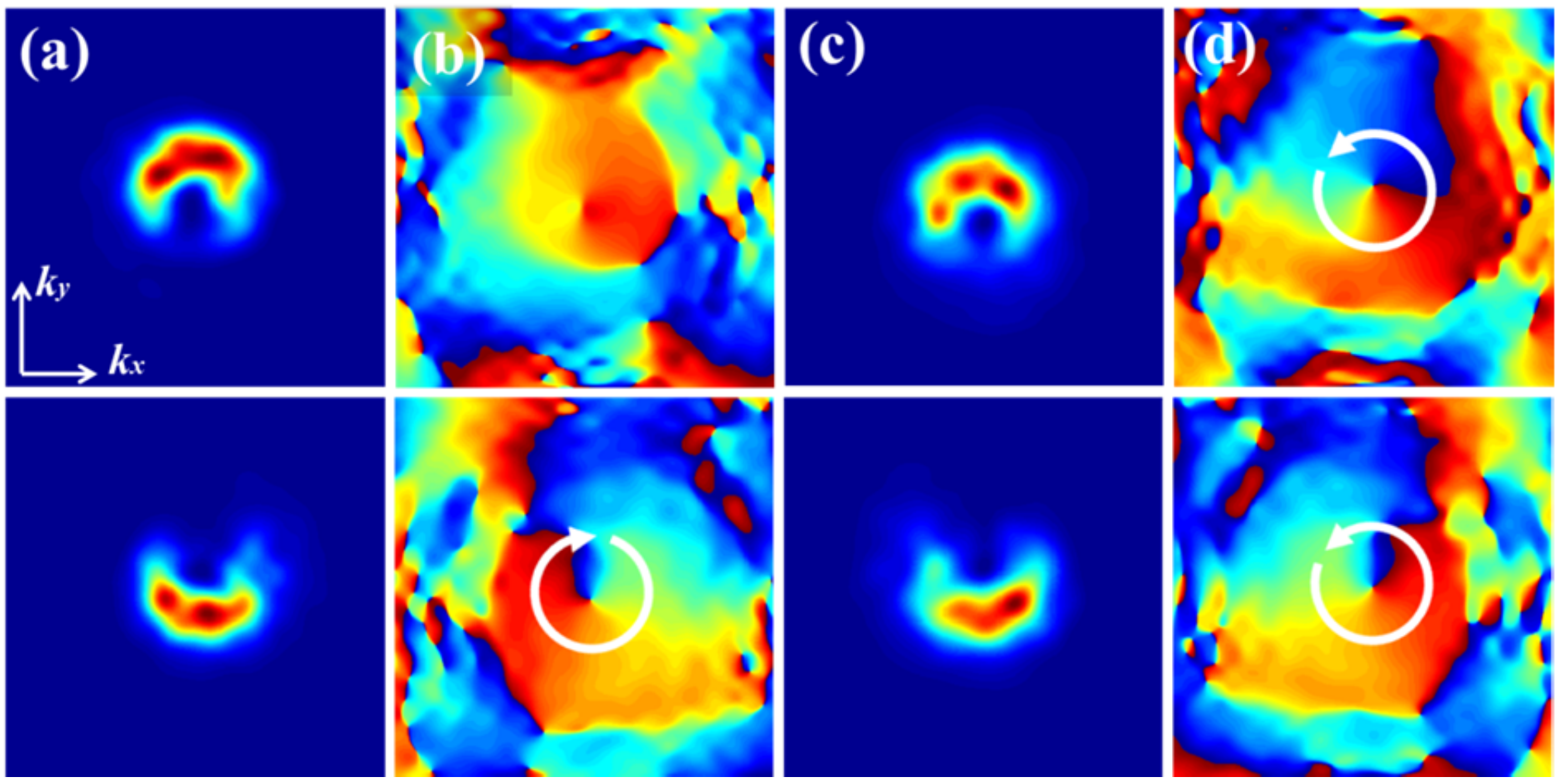

**Fig. 4:** Numerical projection of Bragg-reflected field into two spinor components when two $K$ valleys are selectively excited. Top row corresponds to on-site excitation (Fig. 1 and Fig. 2) for single vortex generation, and bottom row corresponds to inter-site excitation (Fig. 3) for vortex pair generation. (a, b) Intensity and phase patterns of sublattice $\phi_A$ component, (c, d) corresponding results of sublattice $\phi_B$ component. For excitation of two $K$’ valleys, the topological charges for all vortices are reversed.

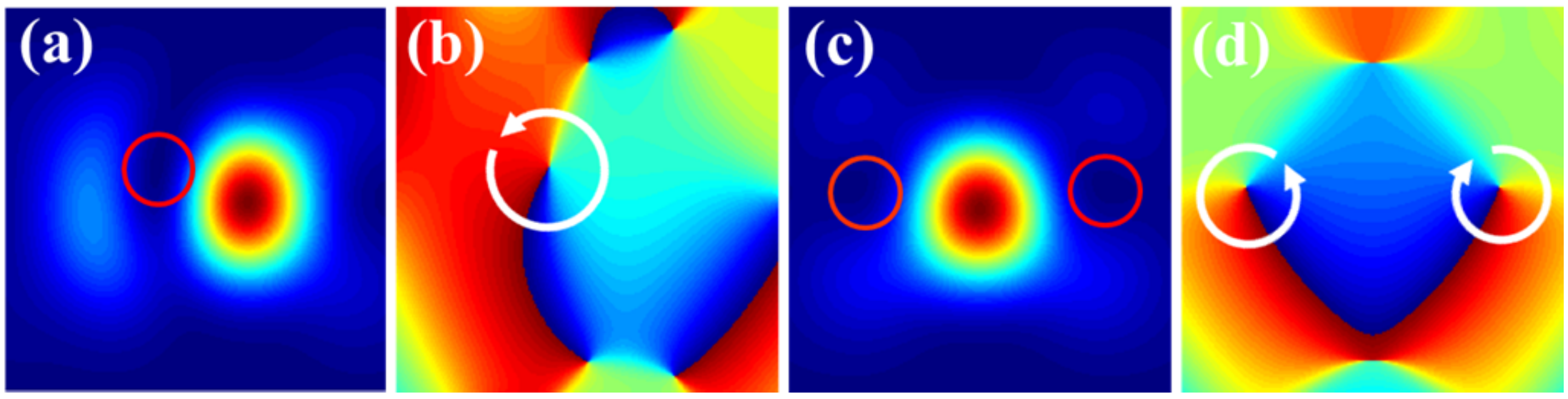

**Fig. 5.** Calculated intensity and phase of the Bragg-reflected field into the third valley from the three band model. (a)-(d) Shown are the results obtained by solving the effective propagation of Eqs. (3, 4) using normalized parameters z = 0.5, V = 1, with a width beam *w* = 0.8. (a, b) For on-site excitation to generate a single vortex (corresponding to Fig. 1, Fig 2). (c, d) For inter-site excitation to generate a vortex-antivortex pair (corresponding to Fig. 3).